\documentclass[twocolumn,           
               showpacs,            
               nopreprintnumbers,     
               aps,                 
               prd,          	    
               letterpaper,             
              groupeaddress,      
               nofootinbib,         
               tightenlines,        
               floats,floatfix,      
               showkeys
               ]{revtex4-1}
               
\usepackage[toc,page]{appendix}
\usepackage{graphicx}
\usepackage{dcolumn}
\usepackage{bm}
\usepackage{amsmath}
\usepackage{amsfonts,amssymb}
\usepackage{soul}
\usepackage{color}

\usepackage{xcolor}
\usepackage{enumerate}
\usepackage{float}
\usepackage{subfigure}
\usepackage{multirow,tabularx, booktabs}
\newcolumntype{C}{>{\centering\arraybackslash}X}

\usepackage{orcidlink}

\begin{document}

\title{Constraints on Tsallis cosmology using recent low and high redshift measurements}

\newcommand{\orcidauthorA}{0000-0003-0405-9344} 
\newcommand{\orcidauthorB}{0000-0003-4666-7890} 
\newcommand{\orcidauthorC}{0000-0003-2293-1802} 
\newcommand{\orcidauthorD}{0000-0002-4276-2906}

\author{M. L. Mendoza-Mart\'inez$^{1}$\orcidlink{\orcidauthorD}}

\author{A. Cervantes-Contreras$^{2}$\orcidlink{\orcidauthorB}}

\author{J. J. Trejo-Alonso$^{2}$\orcidlink{\orcidauthorC}}

\author{A. Hernandez-Almada$^{2}$\orcidlink{\orcidauthorA}}
\email{ahalmada@uaq.mx, corresponding author}

\affiliation{$^1$ Tecnologico de Monterrey, Escuela de Ingenier\'ia y Ciencias, Epigmenio Gonz\'alez 500, San Pablo, 76130, Santiago de Quer\'etaro Qro. M\'exico.}
\affiliation{$^2$ Facultad de Ingenier\'ia, Universidad Aut\'onoma de
Quer\'etaro, Centro Universitario Cerro de las Campanas, 76010, Santiago de 
Quer\'etaro, M\'exico}

\begin{abstract}
Recently Tsallis cosmology has been presented as a novel proposal for alleviating both $H_0$ and $\sigma_8$ tensions. Hence a universe filled with matter and radiation as perfect fluids and considering the Tsallis entropy is confronted using recent cosmological measurements coming from cosmic chronometers, type Ia supernovae, hydrogen II galaxies, quasars, and baryon acoustic oscillations. Following a Bayesian Monte Carlo Markov Chain analysis and combining the samples, we constrain the main characteristic parameters $\alpha = 1.031^{+0.054}_{-0.051}$ and $\delta = 1.005^{+0.001}_{-0.001}$ where the uncertainties are $1\sigma$ confidence level. Additionally, we estimate the current deceleration parameter $q_0=-0.530^{+0.018}_{-0.017}$, the deceleration-acceleration transition redshift $z_T=0.632^{+0.028}_{-0.028}$ and the age of the universe $\tau_U = 12.637^{+0.067}_{-0.066}\,\rm{Gyrs}$ which are in agreement with the standard cosmology ($\Lambda$CDM) within $1.5\sigma$. Furthermore, we find that the dark energy equation of state is consistent with both phantom and quintessence behaviors within $1\sigma$ in the past and converging to $\Lambda$CDM in the future. Additionally, we find agreement values of $H_0$ within $1\sigma$ with the SH0ES values when the CMB distance priors are added to the analysis, suggesting that $H_0$ tension could be alleviated. Finally, Tsallis cosmology is preferred over $\Lambda$CDM by the combined data that motives further studies at the perturbation level.
\end{abstract}
\pacs{cosmology; dark energy; entropy}
\maketitle

\section{Introduction}

Since the discovery of the accelerated expansion of the universe at current epochs by using supernovae type Ia (SNIa) \citep{Riess:1998,Perlmutter:1999} and later confirmed by the satellites WMAP and Planck using the cosmic microwave radiation (CMB) \citep{wmap:2007,Planck:2018}, new open questions about its origin and nature are open up to today. This phenomenon is also namely as dark energy (DE) and the simplest proposal is the well known cosmological constant ($\Lambda$) which is part of the $\Lambda$-Cold Dark Matter ($\Lambda$CDM) model. And is, considered the baseline cosmological model built in General Relativity (GR) framework, which describes both DE and dark matter (DM), the latter modeled as pressureless non-relativistic particles. Based on $\Lambda$CDM, the universe is constituted by about $69\%$ of DE, $26\%$ of DM, and the rest of the visible components (baryons, photons, neutrinos) is about $5\%$ \cite{Planck:2018}.

Although $\Lambda$CDM describes successfully most of the astrophysical and cosmological observations, it presents several problems. Some are related to $\Lambda$ as the fine-tuning and coincidence problems. While the first problem is related to the magnitude order difference between observations and theoretical estimates on the $\Lambda$ value (about 120 orders of magnitude), in the second problem both DE and DM densities are reported with similar values despite their dynamics being different. Recently there has been a tension in the Hubble constant ($H_0$), e.g., in the current rate of universe expansion. This tension consists of disagreement about $5\sigma$ between the $H_0$ value obtained using SNIa (late universe) and the value using CMB (early universe, based on $\Lambda$CDM). Additionally, we can also mention $S_8$ tension which there is a tension of about $3\sigma$ in the quantity $S_8=\sigma_8\sqrt{\Omega_{m0}/0.3}$ where $\sigma_8$ is the amplitude of matter perturbations smoothed over $8h^{-1}\,$Mpc and $\Omega_{m0}$ is the matter density when it is estimated using CMB anisotropies by Planck \cite{Planck:2018} and those from Kilo-Degree Survey (KiDS-1000) \cite{DES:2021wwk,KiDS:2020suj, Heymans_2021}. Some review about $\Lambda$CDM problems and possible solutions see for instance \cite{DiValentino2021,Motta2021,Perivolaropoulos_2022,Hu2023}.

In this way, many theories and models attempt to explain the DE and DM, some from the most fundamental point of view, like modifying the Einstein-Hilbert action through $f(R)$ theories of gravity \cite{sotirio,Shinichi:2011,Nojiri2017} where motivation comes from high-energy physics, cosmology, and astrophysics. Among numerous alternatives to Einstein's theory of gravity, theories that include higher-order curvature invariants. Torsion in gravity $f(T)$, where $T$ is the trace of the stress-energy tensor and teleparallel gravity \citep{telepara}, brings gravity closer to its gauge formulation and incorporates spin in a geometric description. $f(R,T)$ gravity \citep{fdert} are modified theories of gravity, where the gravitational Lagrangian is given by an arbitrary function of the Ricci scalar $R$ and $T$. Unimodular gravity \cite{Carballo,Aspeitia:2021} based on the invariance under a restricted group of diffeomorphisms and Weyl rescalings that leave the determinant of the metric invariant. In the extra dimensions models \cite{ARKANIHAMED}, the gravitons can freely propagate in the new dimensions, at sub-weak energies, the Standard Model fields must be localized to a 4-dimensional manifold of weak scale. Scalar-tensor theories \cite{Coley2003}, gravity theories with non-minimally coupled scalar fields. Some authors have included Thermodynamics in gravity \cite{entropy,padman,Odintsov2023pdu39} theories in which gravity arises as a consequence of entropy. However, we must distinguish between two approaches to this idea: holographic gravity \cite{Nojiri2022,Odintsov2023}, in which Einstein’s equation arises from keeping entropy stationary in equilibrium under variations of the geometry and quantum state of a small region, and thermodynamic gravity, in which Einstein’s equation emerges as a local equation of state from constraints on the area of a dynamical lightsheet in a fixed spacetime background, here the universe is considered filled by DM and DE fluids, applying the first law of thermodynamics to the radius of the observable universe \citep{Zhong-Ying:2015}. In this context, the Tsallis cosmology \citep{Sheykhi:2018}, is an approach that incorporates the entropy into the first law of thermodynamics to produce modified Friedmann cosmological equations, through the Tsallis non-additive entropy \citep{Tsallis:1988} instead of the usual Bekenstein-Hawking one, with interesting results \citep{Jizba:2022,Nojiri:2022,Zamora:2022}, due to the non-additive Tsallis entropy has an exponent $\delta$ that quantifies the departure from standard entropy, and for particular $\delta$ choices can obtain a phantom effective dark energy, which is known to be one of the sufficient mechanisms that could alleviate $H_0$ tension. An increased friction term and an effective Newton's constant smaller than the usual one can even be obtained, and thus the $\sigma_8$ tension could also be solved \cite{Basilakos_2024}. Additionally, this entropy can also affect the production of baryon asymmetry in thermal leptogenesis and reduce the required right-handed neutrino mass scale \cite{Dehpour_2024} and could explain the origin of matter-antimatter asymmetry \cite{Luciano2022-gv}.

Motivated mainly by the recent study \cite{Basilakos_2024} regarding Tsallis cosmology could alleviate both $H_0$ and $\sigma_8$ tensions, this paper aims to test the Tsallis cosmology at the background level using low redshift measurements corresponding to cosmic chronometers, luminosity distance modulus from SNIa, galaxies with $H_{II}$ and quasars, and data from high redshift as Barion Acoustic Oscillations and CMB distance prior.
The paper is divided as follows. In Section \ref{model} a naive derivation of the cosmological model is presented, showing the equations necessary to determine the evolution of the universe, in Section \ref{data} we present a description of the data used in this work, and finally in Section \ref{conclu} the results obtained are present, and provide the conclusions.

\section{Tsallis cosmology}\label{model}

In this section, the model that we will use is briefly described, which was developed by Lymperis $et\, al.$ \cite{lymperis}, where they consider a flat universe ($k=0$) described by the homogeneous and isotropic Friedmann-Lemaitre-Robertson-Walker metric, filled by two perfect fluids. One of them describes the behavior of dark matter and baryons with the equation of state (EoS) $\omega_{m}=p_m/\rho_m=0$, where $p_m$ is the pressure of the matter fluid and $\rho_m$ is the energy density. Additionally, a perfect fluid for relativistic species with EoS $\omega_r=1/3$, being $\rho_r$ the energy density. Furthermore, these fluids satisfy the continuity equations 
\begin{equation} \label{eq:continuity}
    \dot{\rho}_m + 3H\rho_m = 0\,,\quad \quad \dot{\rho}_r + 4H\rho_r = 0\,.
\end{equation}
Additionally, instead of using the traditional Friedmann equation given by GR for these two fluids, {\it i.e.} $H^2=8\pi G/3\sum_i\rho_i$, a modified Friedmann equation is obtained through the first law of thermodynamics by considering the Tsallis entropy, let us remember that the thermodynamics approach does not lead to new gravitational modifications since one needs to consider a specific modified gravity a priori. The Tsallis entropy is defined by \cite{Tsallis:1988,PhysRevLett.80.53,PhysRevLett.84.2770,barboza} 
\begin{equation}\label{tsa}
    S=\frac{\tilde{\alpha}}{4G}A^{\delta}\,
\end{equation}
where $\hbar=k_B=c=1$, $A\propto L^2$, is the area of the system with characteristic length $L$, $\tilde{\alpha}$ is a positive constant with dimensions $[L^{2(1-\delta)}]$, and $\delta$ denotes the non-additivity parameter. As was mentioned before, this entropy is reduced to the Bekenstein-Hawking entropy when $\delta=1$ and $\bar{\alpha}=1$.

We summarize how to obtain the Friedmann equation corresponding to Tsallis entropy as presented in \cite{lymperis}. It is worth mentioning that this procedure is analogous to introducing an extra effective DE fluid into the traditional Friedmann equations.

From $-dE=T\,dS$, with $-dE=A(\rho_t+p_t)H\,\tilde{r}_a\,dt$, where $\rho_t=\rho_m+\rho_r$, $p_t=p_m+p_r$, $A=4\pi{\tilde{r}_a}^2$ is the apparent universe area, $T=1/2\pi\tilde{r}_a$ the apparent universe temperature, and being $\tilde{r}_a=1/H$ the apparent universe radius. Thus the change of the Tsallis entropy Eq. (\ref{tsa}) is given by 
$$dS=(4\pi)^{\delta}\frac{\delta\tilde{\alpha}}{2G}{\tilde{r}_a}^{2\delta-1}\,\dot{\tilde{r}}_a\,dt\,,$$
where $(\,\dot{}\,) = d/dt$ is the derivative with respect to cosmic time, $t$. This latter expression can be expressed as
\begin{equation}\label{eq:Friedmann}
-\frac{(4\pi)^{2-\delta}G}{\tilde{\alpha}}(\rho_t+p_t)=\delta\frac{\dot{H}}{H^{2(\delta-1)}}    \,,
\end{equation}

together with the conservation equations, the Eq. \eqref{eq:continuity}, and by integrating Eq. (\ref{eq:Friedmann}) we get
\begin{equation}\label{eq:acceleration}
\frac{2(4\pi)^{2-\delta}G}{3\tilde{\alpha}}\rho_t=\frac{\delta}{2-\delta}\left(H^2\right)^{2-\delta}-\frac{\tilde{\Lambda}}{3\tilde{\alpha}}\,,    
\end{equation}
where $\tilde{\Lambda}$ is an integration constant which can be considered as the cosmological constant. Here we can remark that the cosmological constant appears naturally in this model instead of including it by hand as in the $\Lambda$CDM model. With the use of Tsallis entropy, the equation has three free parameters $\delta$, $\tilde{\alpha}$, and $\tilde{\Lambda}$.

We can now reorganize the Eqs. \eqref{eq:Friedmann} and \eqref{eq:acceleration} by defining an effective DE perfect fluid and writing the Friedmann equations as
\begin{eqnarray}
H^2&=&\frac{8\pi G}{3}(\rho_m+\rho_r+\rho_{DE})\,,\\
\dot{H}&=&-4\pi G(\rho_m+p_m+\rho_r+p_r+\rho_{DE}+p_{DE})
\end{eqnarray}
where the effective DE, energy density, and pressure are defined as
\begin{eqnarray}
\rho_{DE}&=&\frac{3}{8\pi G}\Bigg[\frac{\Lambda}{3}+H^2\Big(1-\alpha\frac{\delta}{2-\delta}H^{2(1-\delta)}\Big)\Bigg]\,,\\
p_{DE}&=&-\frac{1}{8\pi G}\Bigg[\Lambda+2\dot{H}\Big(1-\alpha\delta H^{2(1-\delta)}\Big) \Bigg.  \nonumber \\
& & \Bigg. +3H^2\Big(1-\alpha\frac{\delta}{2-\delta}H^{2(1-\delta)}\Big)\Bigg]\,,
\end{eqnarray}
where $\Lambda:=(4\pi)^{\delta-1}\tilde{\Lambda}$ and $\alpha:=(4\pi)^{\delta-1}\tilde{\alpha}$.

Additionally, the DE EoS can be obtained by the ratio of these two quantities and give
\begin{equation}
\omega_{DE}=\frac{P_{DE}}{\rho_{DE}}=-1-\frac{2\dot{H}\Big(1-\alpha\delta H^{2(1-\delta)}\Big)}{\Lambda+3H^2\Big(1-\frac{\alpha\delta}{2-\delta}H^{2(1-\delta)}\Big)}	\,.
\end{equation}

Now the first Friedmann equation becomes $\Omega_r+\Omega_m+\Omega_{DE}=1$, where $\Omega_r\equiv\frac{8\pi G}{3H^2}\rho_r$, $\Omega_m=\Omega_{m0}H^2_0/a^3H^2$ (remembering that $\omega_m=0$), and the Hubble parameter can be written as
\begin{equation}\label{hub}
H(a)=\frac{\sqrt{\Omega_{m0}H_0}}{\sqrt{a^3(1-\Omega_{DE}(a)-\Omega_r(a))}}\,,
\end{equation}
where $a=1/(1+z)$ is the scale factor, and $z$ is the redshift variable. Therefore, differentiating Eq. (\ref{hub}) provides
\begin{equation}\label{doh}
\dot{H}=-\frac{H^2}{2}\Bigg[\frac{3\Omega_{m0}+4\Omega_{r0}(1+z)}{\Omega_{m0}+\Omega_{r0}(1+z)} + \frac{(1+z)\Omega'_{DE}}{1-\Omega_{DE}}\Bigg]	\,,
\end{equation}
where the prime, $(^\prime)=d/dz$, denotes derivative with respect to the redshift, and $\Omega'_{DE}(z)$ is written as
\begin{equation}
\Omega'_{DE}(z)=\mathcal{A}(z)\mathcal{B}^{\frac{1}{\delta-2}}(z)\Big[\frac{\mathcal{B}^{-1}(z)\mathcal{C}(z)}{\alpha\delta}-1\Big]\,,	
\end{equation}
where $\mathcal{A}(z)=H_0^2[3\Omega_{m0}(1+z)^2+4\Omega_{r0}(1+z)^3]$, $\mathcal{B}(z)=\frac{2-\delta}{\alpha\delta}\{H_0^2[\Omega_{m0}(1+z)^3+\Omega_{r0}(1+z)^4]+\frac{\Lambda}{3}\}$, and $\mathcal{C}(z)=H_0^2[\Omega_{m0}(1+z)^3+\Omega_{r0}(1+z)^4]$. Finally, by eliminating $\dot{H}$ through Eq. (\ref{doh}), the DE EoS is given as follows
\begin{widetext}
\begin{align}\label{eq:wz}
\omega_{DE}(z)=-1 +\frac{\bigg[\frac{\mathcal{A}(z)(1+z)}{\mathcal{C}(z)}{(1-\Omega_{DE})+(1+z)\Omega'_{DE}\bigg]\Bigg[1-\alpha\delta\bigg(\frac{\mathcal{C}(z)}{1-\Omega_{DE}}\bigg)^{1-\delta}\Bigg]}}	{(1-\Omega_{DE})\Bigg\{\frac{\Lambda(1-\Omega_{DE})}{\mathcal{C}(z)}+3\bigg[1-\frac{\alpha\delta}{2-\delta}\bigg(\frac{\mathcal{C}(z)}{1-\Omega_{DE}}\bigg)^{1-\delta}\bigg]\Bigg\}},
\end{align}
\end{widetext}

and the deceleration parameter $q(z)\equiv-1-\dot{H}/{H^2}$ can be obtained from Eq. (\ref{doh})
\begin{equation}
q(z)=-1+\frac{1}{2}\left[\frac{3\Omega_{m0}+4\Omega_{r0}(1+z)}{\Omega_{m0}+\Omega_{r0}(1+z)}+\frac{(1+z)\Omega'_{DE}}{1-\Omega_{DE}}\right] . 
\end{equation}
In summary, in the constructed modified cosmological scenario, the Eqs. (\ref{hub})-(\ref{eq:wz}) can determine the evolution of the universe. For detailed calculations, see \cite{lymperis} and references therein.

\section{Data}\label{data}
This section presents the dataset used to establish bounds over the Tsallis cosmology.

\begin{itemize}
    \item {\it Cosmic chronometers} (DA). The DA sample are 33 cosmological model independent measurements in the redshift range $0.07<z<1.965$ obtained through the differential age method \cite{Moresco:2016mzx, Moresco:2015cya}. This sample contains two new DA points, $H(z=0.8)=113.1 \pm 15.1\,$km/s/Mpc \cite{Jiao_2023}   and $H(z=1.26) = 135 \pm 65\,$km/s/Mpc \cite{Tomasetti_2023}.
    \item{\it Type Ia Supernovae} (SNIa). A sample of 1701 correlated measurements of SNIa luminosity modulus covers a region $0.001<z<2.26$ in the redshift \cite{Scolnic2018-qf, Brout_2022}. This sample is also named Pantheon+ in the literature.
    \item {\it HII Galaxies} (HIIG). A total of 181 measurements of the distance luminosity from HIIG covering the redshift range $0.01<z<2.6$ \cite{GonzalezMoran2019, Gonzalez-Moran:2021drc}.
    \item {\it Quasars} (QSO). A sample of 120 luminosity measurements coming from intermediate-luminosity quasars covers the redshift range $0.462<z<2.73$ \cite{ShuoQSO:2017}. We use a value $l_m=11.03 \pm 0.25\,$pc for the intrinsic length parameter \cite{ShuoQSO:2017}.
    \item {\it Baryon acoustic oscillations} (BAO). The BAO sample contains 15 uncorrelated points in the redshift region $0.15<z<2.33$ presented in \cite{Alam_2021}. This sample also contains the recent point reported by \cite{descollaboration2024dark} in the effective redshift $z_{eff}=0.85$ of the angular distance to sound horizon scale to $D_M(z_{eff})/rd = 19.51\pm0.41$. For this work we use the drag redshift $z_d = 1089.80 \pm 0.21$ reported by Planck \cite{Planck:2018}.
    \item  {\it Cosmic background radiation} (CMB). The CMB distance prior is a useful tool to extract the effective information of CMB power spectrum without working in the perturbative approach. This information is collected in the acoustic scale $l_A$, and in the shift parameter $R$ and $\Omega_{b0}h^2$.  As a first approximation we use the CMB distance prior given by \cite{Chen2019} based on the CMB Planck 2018 data \cite{Planck:2018} and $\omega$CDM.
\end{itemize}

\section{Constraints and results}
To establish bounds over the Tsallis cosmology parameters $(h,\Omega_{b0},\Omega_{m0}, \alpha, \delta)$ we perform a Bayesian statistical analysis based on Monte Carlo Markov Chain (MCMC) tools using the Emcee module under Python environment \cite{Foreman:2013}. As Emcee authors suggest \cite{Foreman:2013} we use the autocorrelation function to verify the convergence of the chains. A total of 4000 chains with 250 steps are analyzed. To find the maximum of the likelihood function for each data set, we use the following priors: flat priors on $h:\,[0.3,1.0]$,  $\alpha:\,[0,2]$, $\delta:\,[0.5,1.5]$, $\Omega_{b0}h^2:\,[0,0.1]$ and a Gaussian prior on $\Omega_{m0}=0.3116\pm 0.006$ \cite{Planck:2018}. We perform our analysis by combining the samples as CC+SNIa+HIIG+QSO and CC+SNIa+HIIG+QSO+BAO, while the first combination contains only information related to the late universe, the second sample contains a BAO sample which requires information related to the early universe (drag redshift). Due to the use of CMB distance prior being an approximation for slight deviation from $\Lambda$CDM, we will leave the corresponding discussion at the end.

\begin{figure*}
    \centering
    \includegraphics[width=0.9\textwidth]{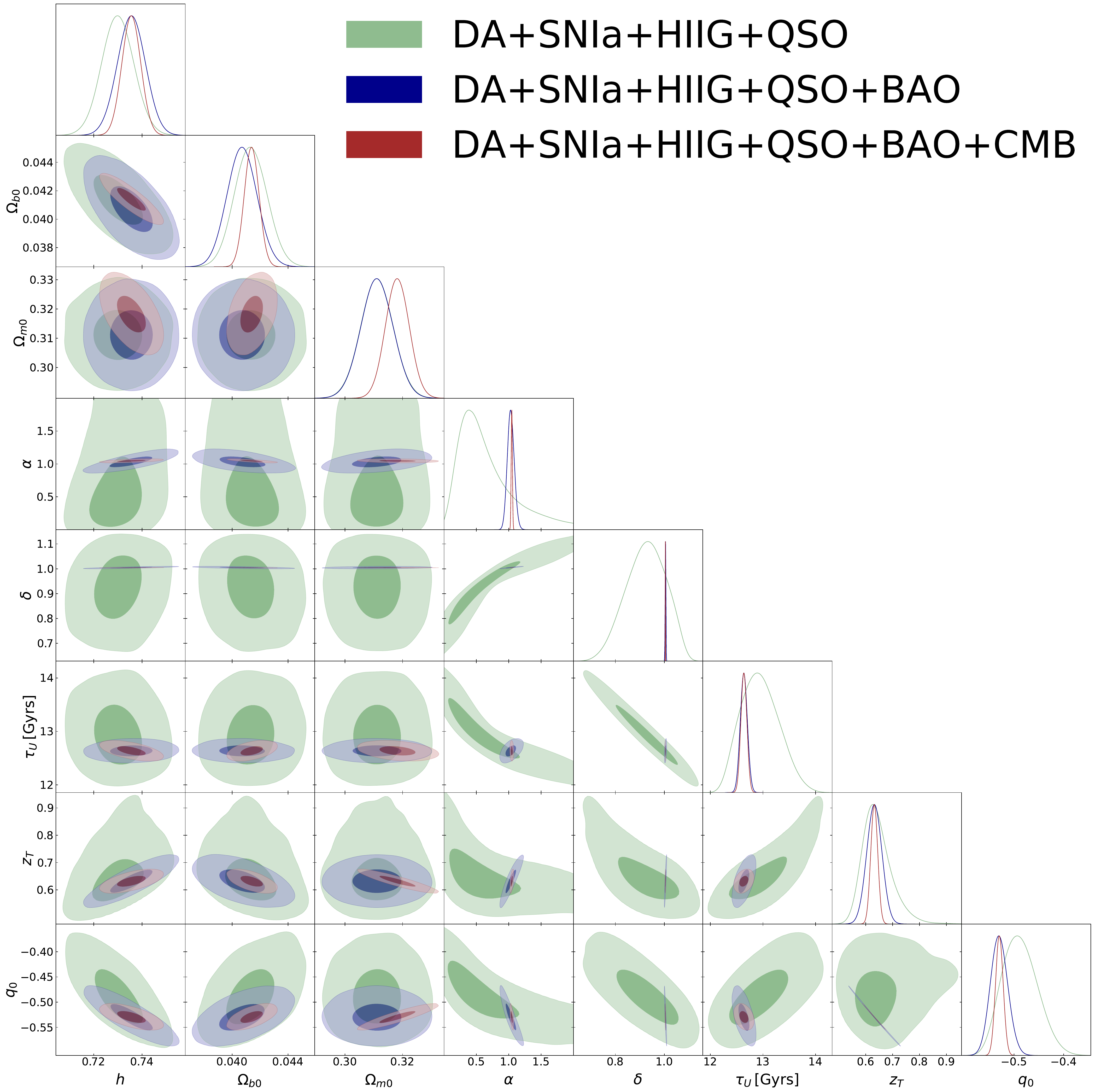}
    \caption{1D posterior distributions and 2D contours at $1\sigma$ (inner region) and $3\sigma$ (outermost region) CL for Tsallis cosmology.}
    \label{fig:contours}
\end{figure*}

The best fit values and their uncertainties at $68\%$ confidence level (CL) for the datasets are reported in Table~\ref{Tab:bf_tsallis} and 1D posterior distributions, and 2D confidence level contours at $1\sigma$ CL and 99.7\% ($3\sigma$) CL are presented in Figure~\ref{fig:contours}. Additionally, the chi-squared function ($\chi^2$) value in the best fit is presented.


\begin{table*}
\begin{tabularx}{0.95\textwidth}{@{} lCCC @{}}
 \toprule
Parameter	& DA+SNIa+HIIG+QSO	& DA+SNIa+HIIG+QSO+BAO & DA+SNIa+HIIG+QSO+BAO+CMB\\
\hline
   \multicolumn{4}{c}{Tsallis cosmology} \\ [0.9ex]
 $h$            & 	$0.730^{+0.007}_{-0.007}$  & 	$0.736^{+0.006}_{-0.006}$  & 	$0.736^{+0.004}_{-0.004}$  \\ [0.9ex] 
 $\Omega_{b0}$  & 	$0.041^{+0.001}_{-0.001}$  & 	$0.041^{+0.001}_{-0.001}$  & 	$0.041^{+0.001}_{-0.001}$  \\ [0.9ex] 
 $\Omega_{m0}$  & 	$0.311^{+0.006}_{-0.006}$  & 	$0.311^{+0.006}_{-0.006}$  & 	$0.318^{+0.004}_{-0.004}$  \\ [0.9ex] 
 $\alpha$       & 	$0.532^{+0.521}_{-0.271}$  & 	$1.031^{+0.054}_{-0.051}$  & 	$1.047^{+0.009}_{-0.009}$  \\ [0.9ex] 
 $\delta$       & 	$0.928^{+0.083}_{-0.088}$  & 	$1.005^{+0.001}_{-0.001}$  & 	$1.004^{+0.0005}_{-0.0005}$  \\ [0.9ex] 
 $\tau_U \,[\rm{Gyrs}]$  & 	$12.925^{+0.392}_{-0.356}$  & 	$12.637^{+0.067}_{-0.066}$  & 	$12.637^{+0.057}_{-0.056}$  \\ [0.9ex] 
 $z_T $         & 	$0.636^{+0.059}_{-0.045}$  & 	$0.632^{+0.028}_{-0.028}$  & 	$0.632^{+0.013}_{-0.013}$  \\ [0.9ex] 
 $q_0 $         & 	$-0.490^{+0.037}_{-0.034}$  & 	$-0.530^{+0.018}_{-0.017}$  & 	$-0.529^{+0.008}_{-0.008}$  \\ [0.9ex] 
$\chi^2$        & $5610.17$  & $5630.55$  & $5635.91$  \\ [0.9ex]    
\hline
\multicolumn{4}{c}{$\Lambda$CDM} \\ [0.9ex]
 $h$  & 	$0.736^{+0.004}_{-0.004}$  & 	$0.726^{+0.003}_{-0.003}$  & 	$0.698^{+0.003}_{-0.003}$  \\ [0.9ex] 
 $\Omega_{b0}$  & 	$0.041^{+0.001}_{-0.001}$  & 	$0.044^{+0.001}_{-0.001}$  & 	$0.047^{+0.0003}_{-0.0003}$  \\ [0.9ex] 
 $\Omega_{m0}$  & 	$0.313^{+0.005}_{-0.005}$  & 	$0.325^{+0.005}_{-0.005}$  & 	$0.292^{+0.004}_{-0.004}$  \\ [0.9ex] 
 $\tau_U \,[\rm{Gyrs}]$  & 	$12.661^{+0.057}_{-0.056}$  & 	$12.712^{+0.057}_{-0.056}$  & 	$13.609^{+0.015}_{-0.015}$  \\ [0.9ex] 
 $z_T $  & 	$0.636^{+0.013}_{-0.013}$  & 	$0.608^{+0.011}_{-0.011}$  & 	$0.691^{+0.010}_{-0.010}$  \\ [0.9ex] 
 $q_0 $  & 	$-0.530^{+0.008}_{-0.008}$  & 	$-0.513^{+0.007}_{-0.007}$  & 	$-0.562^{+0.005}_{-0.005}$  \\ [0.9ex] 
$\chi^2$  & $5612.89$  & $5686.98$  & $5923.72$  \\ [0.9ex] 
\hline
\bottomrule
\end{tabularx}
\caption{Best fit values of the parameters for Tsallis cosmology and $\Lambda$CDM for the dataset used. The last row corresponds to $\chi^2$ value for best fit.}
\label{Tab:bf_tsallis}
\end{table*}

A statistical comparison between the Tsallis model and $\Lambda$CDM is performed by estimating the Akaike information criterion (AIC) \cite{AIC:1974, Sugiura:1978} defined as AIC$\equiv\chi^2 + 2k$, where $k$ is the number of free parameters and $N$ is the total size of the dataset. In this criterion, the model with the lowest AIC value is the one preferred by the data as the following rules. If the difference in AIC value between a given model and the best one ($\Delta$AIC) is less than $4$, both models are equally supported by the data. For $\Delta$AIC values in the interval $4<\Delta$AIC$<10$, the data still support the given model but less than the preferred one. For $\Delta$AIC$>10$, the observations do not support the given model. According to our results presented in Table~\ref{Tab:AICBIC}, we find that both models are equally supported by DA+SNIa+HIIG+QSO but the Tsallis model is preferred by DA+SNIa+HIIG+QSO+BAO.

As a complement, we also estimate the Bayesian Information criterion (BIC) \cite{schwarz1978} defined as BIC$\equiv\chi^2+k\log(N)$, which gives a stronger penalty than AIC due to the number of degree of freedom. This criterion is interpreted as evidence against a candidate model being the best model (also the best model is the one with a lower BIC value). BIC discriminates according to the following rules. For $\Delta$BIC$<2$, there is no appreciable evidence against the model. If $2<\Delta$BIC$<6$, there is modest evidence against the candidate model. For the interval $6<\Delta$BIC$<10$, the evidence against the candidate model is strong, and even stronger evidence against is when $\Delta$BIC$>10$. Based on Table~\ref{Tab:AICBIC}, we have stronger evidence against Tsallis cosmology for DA+SNIa+HIIG+QSO but when the full dataset is considered the stronger evidence against is for $\Lambda$CDM. For both criteria, negative values on the $\Delta$AIC or $\Delta$BIC mean a disfavoured result over $\Lambda$CDM.

\begin{table*}
\begin{tabularx}{0.8\textwidth}{@{} lCCC @{}}
\hline
   Dataset               &    AIC(Tsallis)   &  AIC($\Lambda$CDM) & $\Delta$AIC \\
\hline
DA+SNIa+HIIG+QSO         & 5620.17	         &  5618.89            &    1.28  \\ [0.9ex] 
DA+SNIa+HIIG+QSO+BAO     & 5640.55	         &  5692.98            &  -52.43  \\ [0.9ex]  
DA+SNIa+HIIG+QSO+BAO+CMB & 5645.91	         &  5929.72            & -283.81  \\ [0.9ex] 

\hline
                         & BIC(Tsallis)      &  BIC($\Lambda$CDM) & $\Delta$BIC \\ [0.9ex] 
DA+SNIa+HIIG+QSO         &	5648.26	         &  5635.74           &	  12.51  \\ [0.9ex] 
DA+SNIa+HIIG+QSO+BAO     &	5668.67	         &  5709.85           &	 -41.18  \\ [0.9ex] 
DA+SNIa+HIIG+QSO+BAO+CMB &  5674.04	         &  5946.60	          & -272.56  \\ [0.9ex] 

	\hline
 \bottomrule
\end{tabularx}
\caption{Statistical comparison between Tsallis cosmology and $\Lambda$CDM  using AIC and BIC. The quantity $\Delta$ is the difference between the AIC (BIC) values for Tsallis and those for $\Lambda$CDM. Values $\Delta$AIC$<-10$  suggest not support
for $\Lambda$CDM, and values $\Delta$BIC$<-10$ a stronger evidence against it.}
\label{Tab:AICBIC}
\end{table*}

\begin{figure*}
    \centering
    \includegraphics[width=0.32\textwidth]{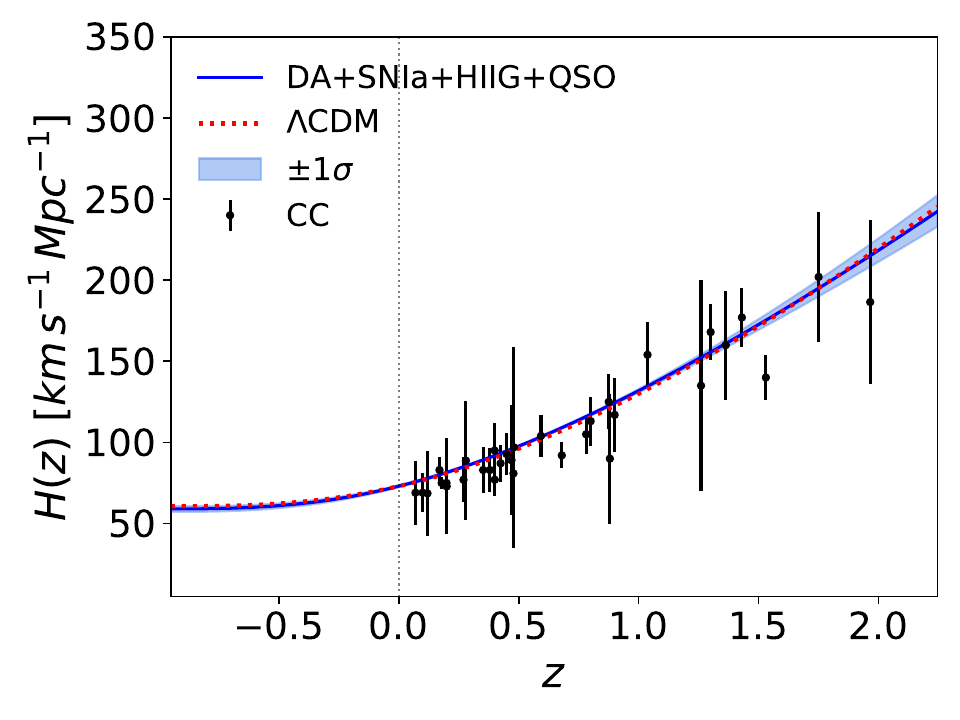}
    \includegraphics[width=0.32\textwidth]{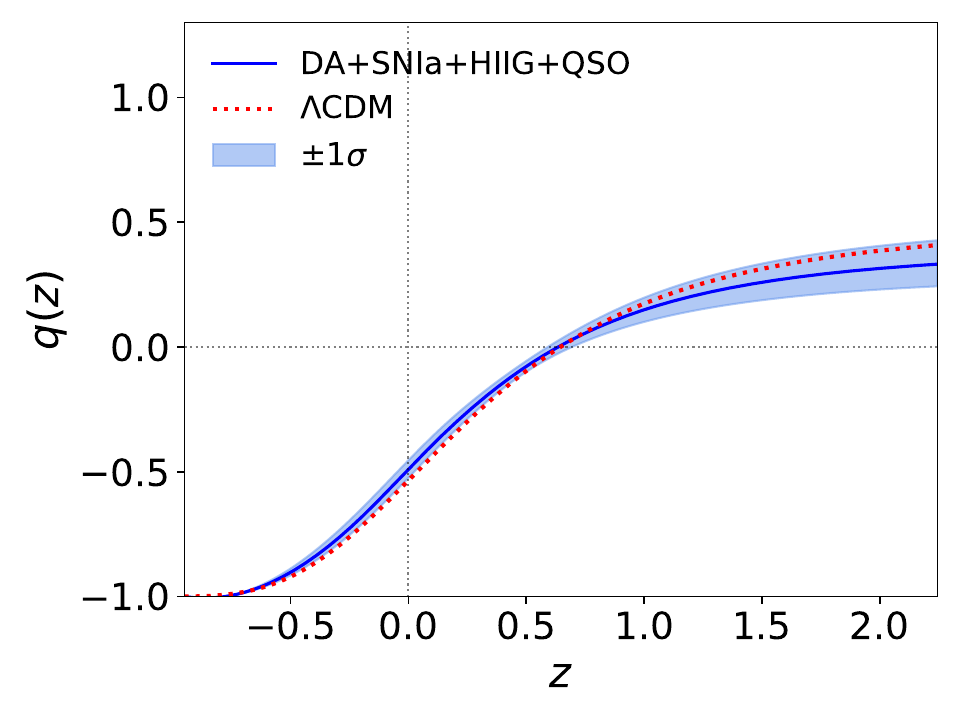}
    \includegraphics[width=0.32\textwidth]{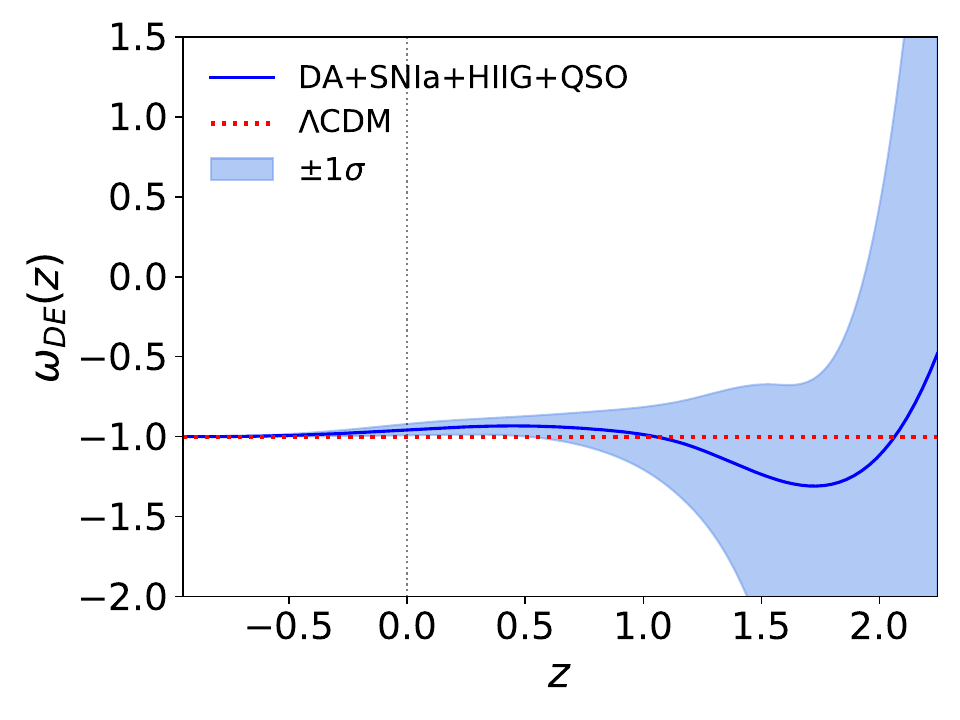}\\
    \includegraphics[width=0.32\textwidth]{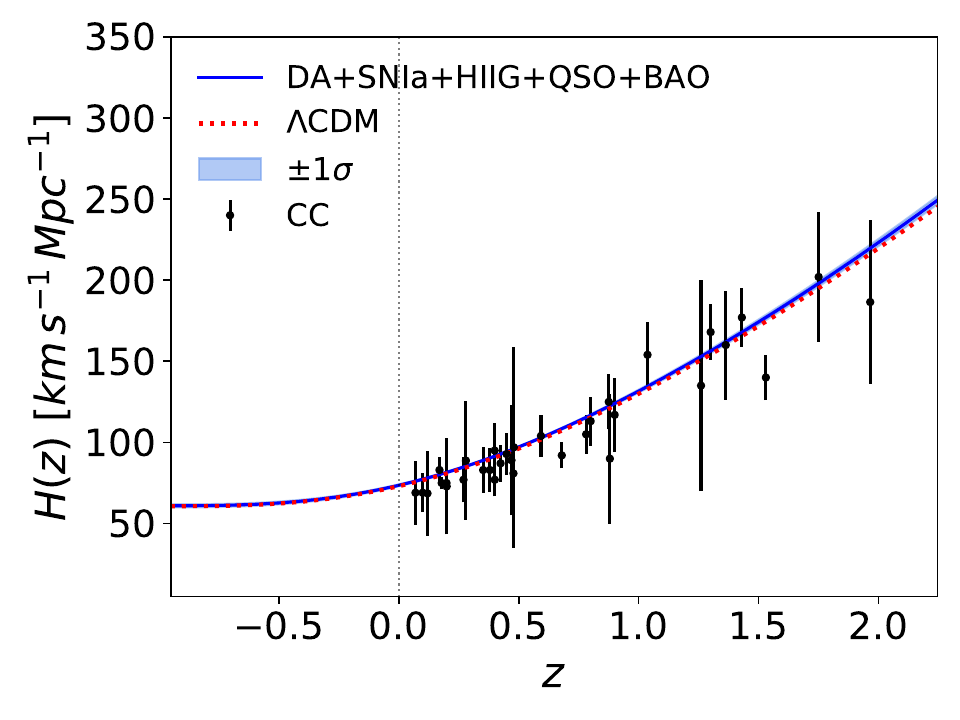}
    \includegraphics[width=0.32\textwidth]{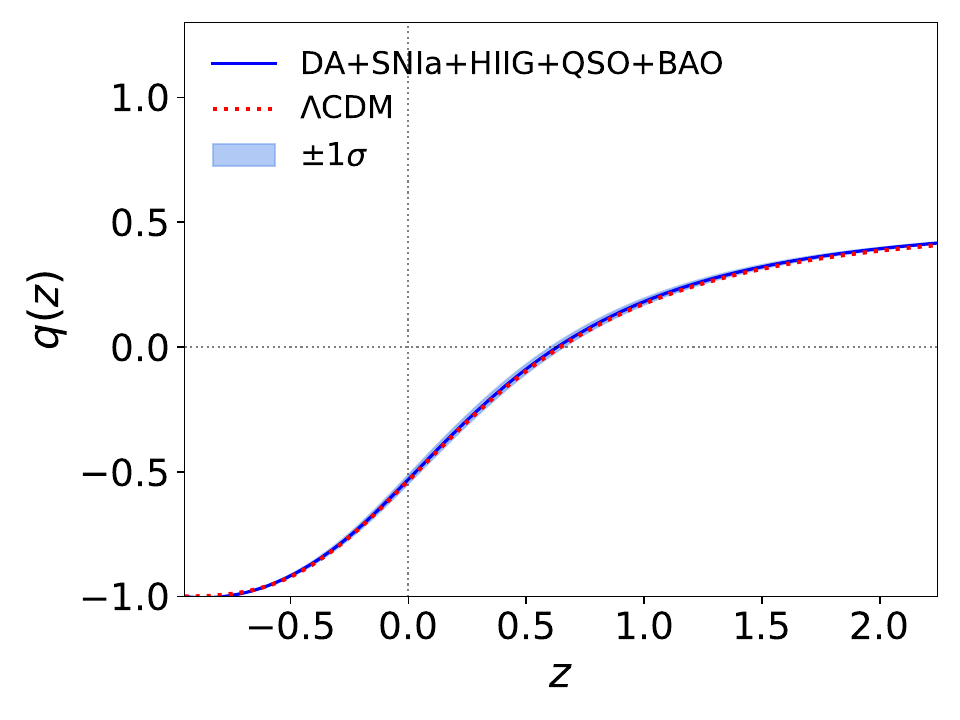}
    \includegraphics[width=0.32\textwidth]{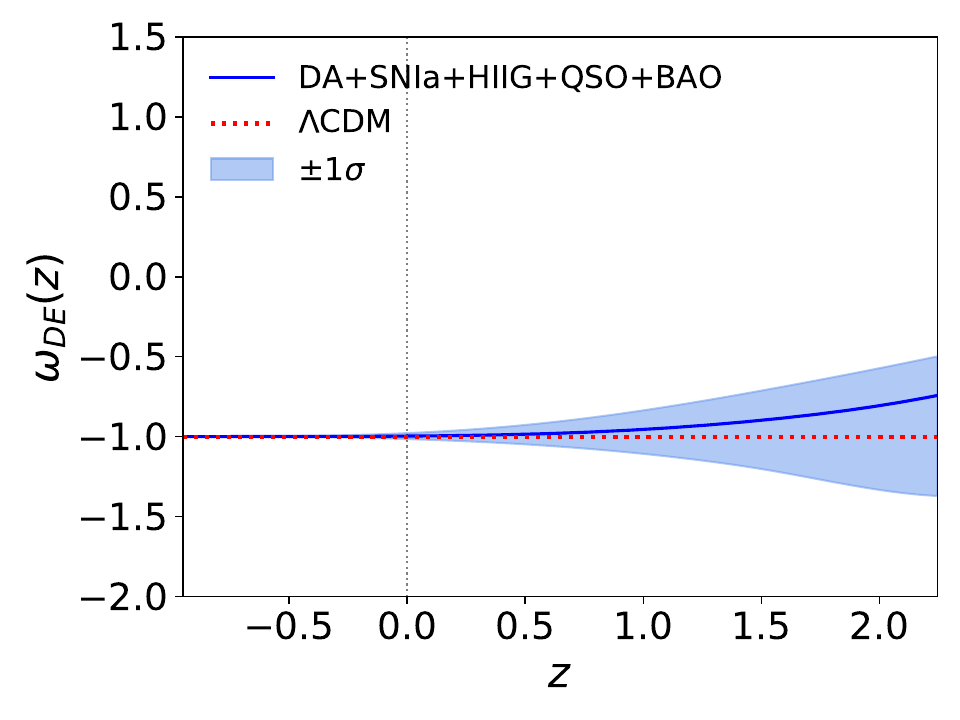}\\
    \includegraphics[width=0.32\textwidth]{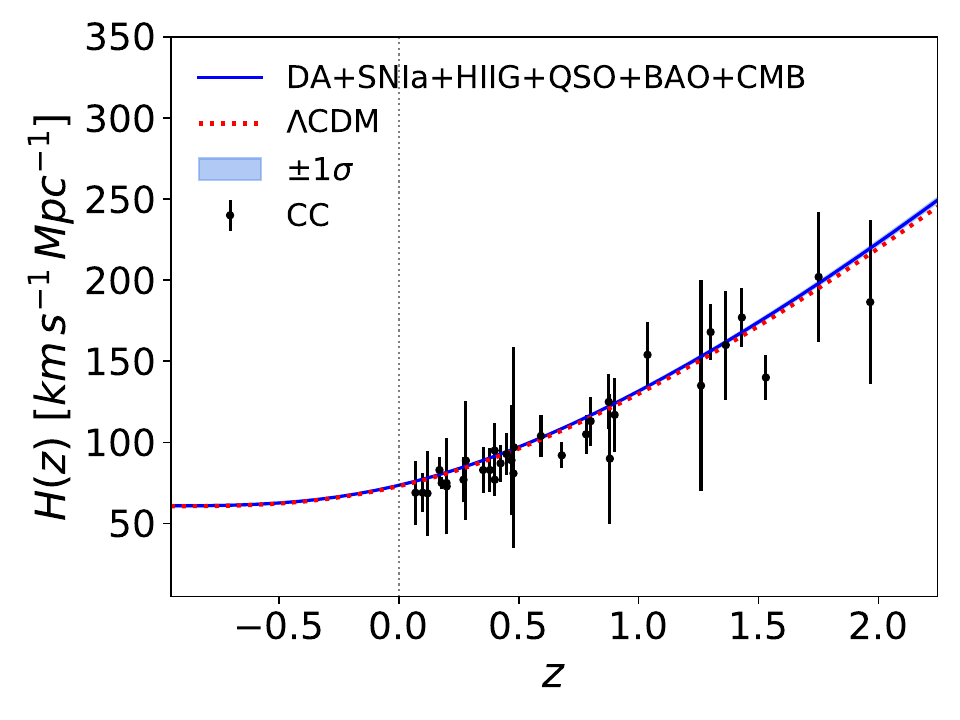}
    \includegraphics[width=0.32\textwidth]{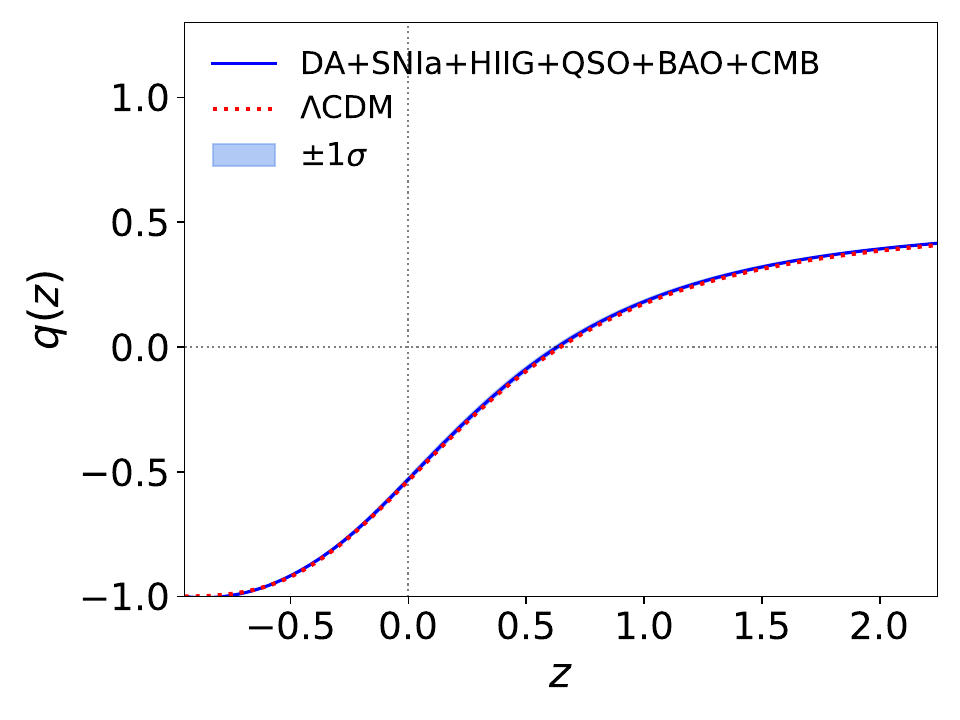}
    \includegraphics[width=0.32\textwidth]{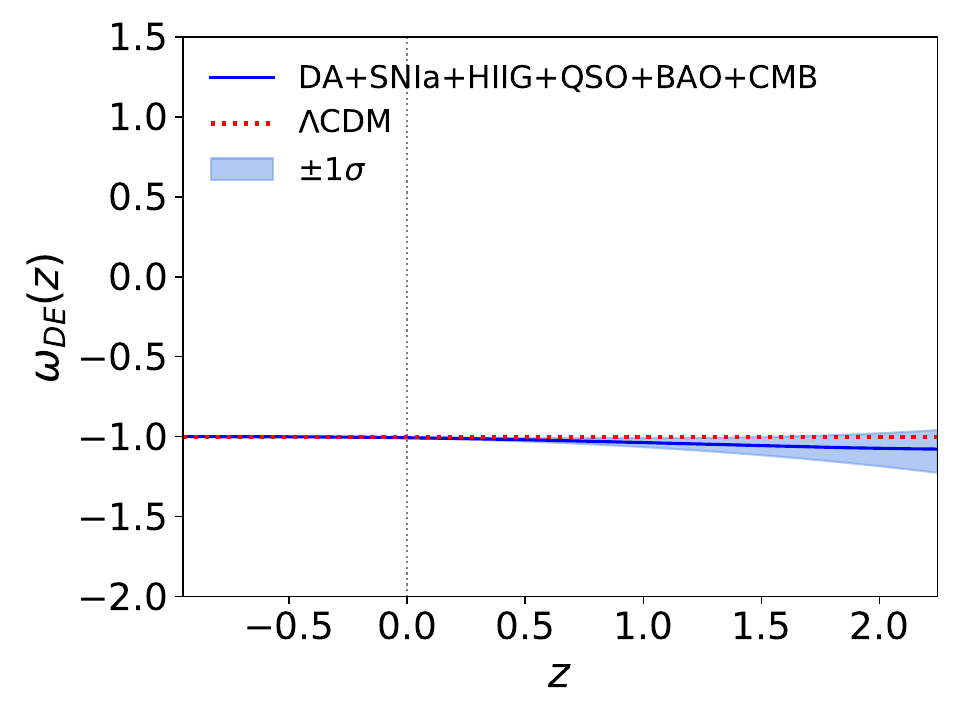}

    \caption{Reconstruction of the Hubble parameter (left panel), deceleration parameter (middle panel), and DE EoS (right panel) for Tsallis cosmology in the region $-0.95<z<2.2$ using DA+SNIa+HIIG+QSO (top panel), DA+SNIa+HIIG+QSO+BAO (middle panel)  and DA+SNIa+HIIG+QSO+BAO+CMB (bottom panel) datasets. The bands represent the uncertainty at $1\sigma$, solid line is the best fit and dotted line is $\Lambda$CDM.}
    \label{fig:cosmo_reco}
\end{figure*}

On the other hand, Figure~\ref{fig:cosmo_reco} displays the reconstruction of the Hubble parameter, the deceleration parameter, and the DE equation of state in the redshift range $-0.95<z<2.2$, for Tsallis cosmology using our results for both datasets. The bands represent their uncertainties at $1\sigma$ around the best fit line (solid line). The corresponding $\Lambda$CDM curves are in dotted lines. We have an agreement within $2\sigma$ for the three mentioned functions. Based on results presented in Table~\ref{Tab:bf_tsallis}, the constrained value of the Hubble constant for Tsallis deviates $0.75\sigma$ ($1.5\sigma$) from the $\Lambda$CDM using DA+SNIa+HIIG+QSO (DA+SNIa+HIIG+QSO+BAO) and the current value ($z=0$) of the deceleration parameter $q_0$ is within $1.5\sigma$ ($0.88\sigma$). Additionally we estimate a deceleration-acceleration transition resdhift $z_T=0.636^{+0.059}_{-0.045}$ and $z_T=0.632^{+0.028}_{-0.028}$ for DA+SNIa+HIIG+QSO and DA+SNIa+HIIG+QSO+BAO respectively, which are in agreement within $1\sigma$ with those $\Lambda$CDM values. Additionally, we also find the age of the universe in agreement within $1\sigma$ for both cosmological models. The right panel of Figure~\ref{fig:cosmo_reco} is the DE EoS for both samples. Although both reconstructions are in agreement with a CC EoS within $1\sigma$ and converging to $w\to-1$ in the future, we observe phantom dynamics in the redshift region $1\lesssim z \lesssim 2$ and $z\lesssim1$ a quintaessence behavior for DA+SNIa+HIIG+QSO dataset. For DA+SNIa+HIIG+QSO+BAO the bestfit curve only lives in the quintaessence region. These results are consistent with those explored by \cite{lymperis}.

Regarding the characteristic parameters of the Tsallis model, we find that Tsallis entropy is in agreement within $1\sigma$ to the limit of the black hole entropy (Bekenstein-Hawking area law entropy) \cite{HAWKING_1974,Hawking1975} when $\alpha=1$ and $\delta=1$ for DA+SNIa+HIIG+QSO data. However, we estimate a deviation of $5\sigma$ in the non-extensive parameter to $\delta=1$ for our stronger constraint. On the other hand, authors in \cite{Asghari_2021} report a constraint over this parameter as $\delta\sim 0.9997$ using six combined likelihoods corresponding to CMB, SNIa, and BAO (see \cite{Asghari_2021} for more details), which is deviated around $3.5\sigma$ from our stronger bound. It is worth mentioning that these results could be in tension with the ones obtained in \cite{Jizba2023}, which establish bounds on $\delta$ in the region $\delta\sim1.499$, which is obtained by taking into account the Big Bang nucleosynthesis and the relic abundance of cold dark matter particles, and is consistent with the formation of primordial light elements, such as deuterium $^2H$ and helium $^4He$.

Finally, the constraints when CMB is added to the samples are discussed. We confirm that Tsallis cosmology could alleviate the $H_0$ tension, we find in agreement the value of $h$ within $1\sigma$ with the one obtained by SH0ES \cite{Riess_2022}. As Tab. \ref{Tab:bf_tsallis} shows, the value of $h$ is moved to the value obtained by Planck value. Tsallis model is preferred over $\Lambda$CDM by the combined sample DA+SNIa+HIIG+QSO+BAO+CMB according to AIC and BIC. As is expected stronger bounds on the reconstruction of the Hubble parameter, deceleration parameter, and the DE EoS are obtained, see Fig. \ref{fig:cosmo_reco}. Furthermore, the value of $q_0$ and $z_T$ are in agreement with those obtained using only early cosmological data and also when BAO is added. As last, the best fit reconstruction of the DE EoS is preferred to be in the phantom region instead of the quintessence for DA+SNIa+HIIG+QSO+BAO data.

\section{Conclusions}\label{conclu}

The goal of the current manuscript was to confront a cosmological model containing two perfect fluids (matter and relativistic species) based on the Tsallis entropy with several cosmological observations. The Friedmann equations presented in Eqs. \eqref{eq:Friedmann} and \eqref{eq:acceleration} were obtained by using a thermodynamics approach (for details in this methodology see \cite{lymperis}). Based on a Bayesian MCMC analysis, the free parameters of the Tsallis model ($h, \Omega_{m}, \alpha, \delta$) were constrained using CC, SNIa, HIIG, QSO, and BAO data using the combination CC+SNIa+HIIG+QSO (our baseline sample)  which corresponds to measurements at low redshift or late time of the universe. A stronger constraint on the free parameters is obtained by adding a BAO sample to our baseline data. It is worth mentioning that BAO dataset contains information related to the early epochs of the universe at the drag redshift. The results were shown in Table \ref{Tab:bf_tsallis} and we found that Tsallis entropy is consistent with the Bekenstein-Hawking entropy within $1\sigma$ for our baseline data and $5\sigma$ of deviation between both entropies for baseline+BAO sample. Also, our stronger constraint results are deviated $3.5\sigma$ with those estimated by \cite{Asghari_2021}  using data from CMB, SNIa, and BAO and are in tension with those results obtained using Big Bang nucleosynthesis and the relic abundance of cold dark matter particles. Furthermore, we also estimate physical parameters as deceleration-acceleration transition redshift $z_T$, deceleration parameter at today $q_0$, age of universe $\tau_U$ that were consistent within up to $1.5\sigma$ with those values for $\Lambda$CDM. Additionally, we find consistent DE EoS behavior with the result obtained by \cite{lymperis} that covers both phantom and quintessence dynamics within $1\sigma$. 

On the other hand, we compare statistically the Tsallis cosmology with the standard cosmology using AIC and BIC. Both criteria suggest that Tsallis cosmology is preferred by baseline+BAO data but a stronger evidence against Tsallis model when the baseline data is only considered.

Regarding to our results when CMB data are added, we found that $H_0$ value is close to the SH0ES value \cite{Riess_2022} within $1\sigma$ while the value from $\Lambda$CDM is moved to the Planck value \cite{Planck:2018}. This result confirms that the Tsallis model could alleviate the $H_0$ tension as was proposed in \cite{Basilakos_2024}. Both cosmographic reconstructions $H(z)$ and $q(z)$ are in agreement with $\Lambda$CDM reconstructions and the value of $z_T$ is deviated $3.6\sigma$. In contrast with the previous results, by adding CMB data, the best fit reconstruction of the DE EoS is moved to the phantom region. Additionally Tsallis cosmology is preferred over $\Lambda$CDM when CMB data is added to baseline+BAO sample based on AIC and BIC but we must be remark that the use of CMB distance prior could give a bias in this estimation.

Finally, we find that Tsallis cosmology is an interesting alternative to $\Lambda$CDM which naturally could solve some of the $\Lambda$ problems as it is its origin. Additionally authors \cite{Basilakos_2024} also discuss how Tsallis entropy could alleviate both $H_0$ and $\sigma_8$ tensions and motivate a perturbative analysis by considering full CMB data and kiDS data that is out of the scope of this work.

\begin{acknowledgments}
The authors thank the anonymous referee for thoughtful remarks and suggestions. Authors thank O. Cornejo-P\'erez for his fruitful discussions. A.H.A. thanks to the support from Luis Aguilar, 
Alejandro de Le\'on, Carlos Flores, and Jair Garc\'ia of the Laboratorio 
Nacional de Visualizaci\'on Cient\'ifica Avanzada. M.L.M.M. thanks to the SNII-Me\'xico, CONAHCyT for support. 
\end{acknowledgments}

\bibliography{main}

\end{document}